\documentclass[namedreferences]{SolarPhysics}
\usepackage[optionalrh,solaenum]{spr-sola-addons} % For Solar Physics
\usepackage{graphicx}        % For eps figures, newer & more powerfull
\usepackage{url}             % For breaking URLs easily trough lines
\ifx \doiurl    \undefined
\def\doiurl#1{\href{http://dx.doi.org/#1}{\textsf{#1}}}\fi \ifx\adsurl    \undefined \def
\adsurl#1{\href{http://adsabs.harvard.edu/abs/#1}{\textsf{#1}}}\fi
\ifx \arxivurl  \undefined \def
\arxivurl#1{\href{http://arxiv.org/abs/#1}{\textsf{#1}}}\fi
\begin{document}

\begin{article}

\begin{opening}

\title{Near-Limb Zeeman and Hanle Diagnostics}

\author{I.S.~\surname{Kim}\sep
        I.V.~\surname{Alexeeva}\sep
        O.I.~\surname{Bugaenko}\sep
        V.V.~\surname{Popov}\sep
        E.Z.~\surname{Suyunova}
       }
\runningauthor{I.S. Kim \textit{et al.}} \runningtitle{Near-Limb
Zeeman and Hanle Diagnostics}

   \institute{Moscow M.V. Lomonosov State University, Sternberg Astronomical
   Institute, Moscow, Russia,
                       email: \href{mailto:kim@sai.msu.ru}{kim@sai.msu.ru}
             }

\begin{abstract}
``Weak'' magnetic-field diagnostics in faint objects near the
bright solar disk are discussed in terms of  the level of
non-object signatures, in particular, on the stray light in
telescopes. Calculated dependencies of the stray light caused by
diffraction at the 0.5, 1.6, and 4 meter entrance aperture are
presented. The requirements for micro-roughness of refractive and
reflective primary optics are compared. Several methods for
reducing the stray light (the Lyot coronagraphic technique,
multiple stages of apodizing in the focal and exit pupil planes,
apodizing in the entrance aperture plane with a special mask) and
reducing the random and systematic errors are noted. An acceptable
level of stray light in telescopes is estimated for the
$V$-profile recording with a signal-to-noise ratio greater than
three. Prospects for the limb chromosphere magnetic measurements
are indicated.
\end{abstract}
\end{opening}

\section{Introduction}
     \label{S-Introduction}
Near-limb Zeeman and Hanle diagnostics are connected with weak
magnetic-field measurements in the upper solar atmosphere:
prominences, the chromosphere, and the corona. Key items of
magnetic measurements in the upper solar atmosphere are
low-scattered-light feed optics (telescopes), an advanced
analyzing equipment (polarimeters), and advanced recording
equipment. So far, such measurements have not become routine in
spite of available advanced coronagraphs, polarimeters, and
recording systems \cite{lin:2004,tomczyk:2008}. This is a task for
forthcoming exciting ground- and space-based projects
\cite{keil:2003,rimmele:2010,wagner:2010,tomczyk:2011,peter:2012}.
Non-solar object signatures in the final focal plane, in
particular, the stray light [$I_{\rm stray}$], the sky brightness
[$I_{\rm sky}$], and the continuum corona [$I_{\rm cont}$]
complicate both linear and circular non-eclipse coronagraphic and
eclipse polarimetry.

In this article we consider general expressions applicable in the
upper solar atmosphere and several previous long-term magnetic
measurements in prominences with the emphasis on $I_{\rm stray}$
of feeding optics (Section 1), dependencies of $I_{\rm stray}$ on
distance caused by diffraction at the edge of an entrance aperture
of 0.5, 1.6, and 4.0 meters and several ways of reducing $I_{\rm
stray}$ (Section 2), reducing random and systematic errors
(Section 3) and acceptable level of $I_{\rm stray}$ for the
$V$-profile recording with a signal-to-noise ratio $>$ three
(Section 4). Finally, prospects for magnetic measurements in the
upper solar atmosphere are noted.

\subsection{General Expressions}
The term ``weak magnetic field'' is used when the Zeeman splitting
[$\delta\lambda_{B}$] is three\,--\,four orders of magnitude less
than the line width [$\Delta\lambda$]: the full width at half
maximum].
$\delta\lambda_{B}=4.67\times10^{-13}g\lambda_{0}^{2}B_{\prime\prime}$,
where $\lambda_{0}$ is the wavelength in \r{A}, $B_{\prime\prime}$
denotes the strength of  longitudinal magnetic field in G, and $g$
is the Land\'e factor. As a rule, the effective Land\'e factor
[$g_{e}$] is used to take into account the different contributions
to the magnetic splitting caused by different components of the
line. The left part of Figure 1 shows $\delta\lambda_{B}$
\textit{versus} $B_{\prime\prime}$ for the chromospheric
He~{\footnotesize I} D{\footnotesize 3} ($g_{e} = 1.12$) and
1083.0 nm ($g_{e}= 1.46$), H {\footnotesize I} H$\alpha$ ($g_{e} =
1.05$),  and coronal Fe {\footnotesize XIII} 1074.7 nm ($g_{e} =
1.46$) lines. The $g_{e}$ used for the He~{\footnotesize I} lines
does not take into account the low-intensity component, that is
valid when $\Delta\lambda_{D3}\ge 0.4$ \r{A}  and
$\Delta\lambda_{1083}\ge1$ \r{A}. $\delta\lambda_{B}$ of the IR
lines differ by a factor of $0.99$ and are presented by the same
curve. The $\delta\lambda_{B}$ range is
$2\times(10^{-4}$\,--\,$10^{-3})$ for $B_{\prime\prime}=10-20$ G.

\begin{figure}    %%%%%%%%%%%%%%%%%% FIGURE 1
   \centerline{\includegraphics{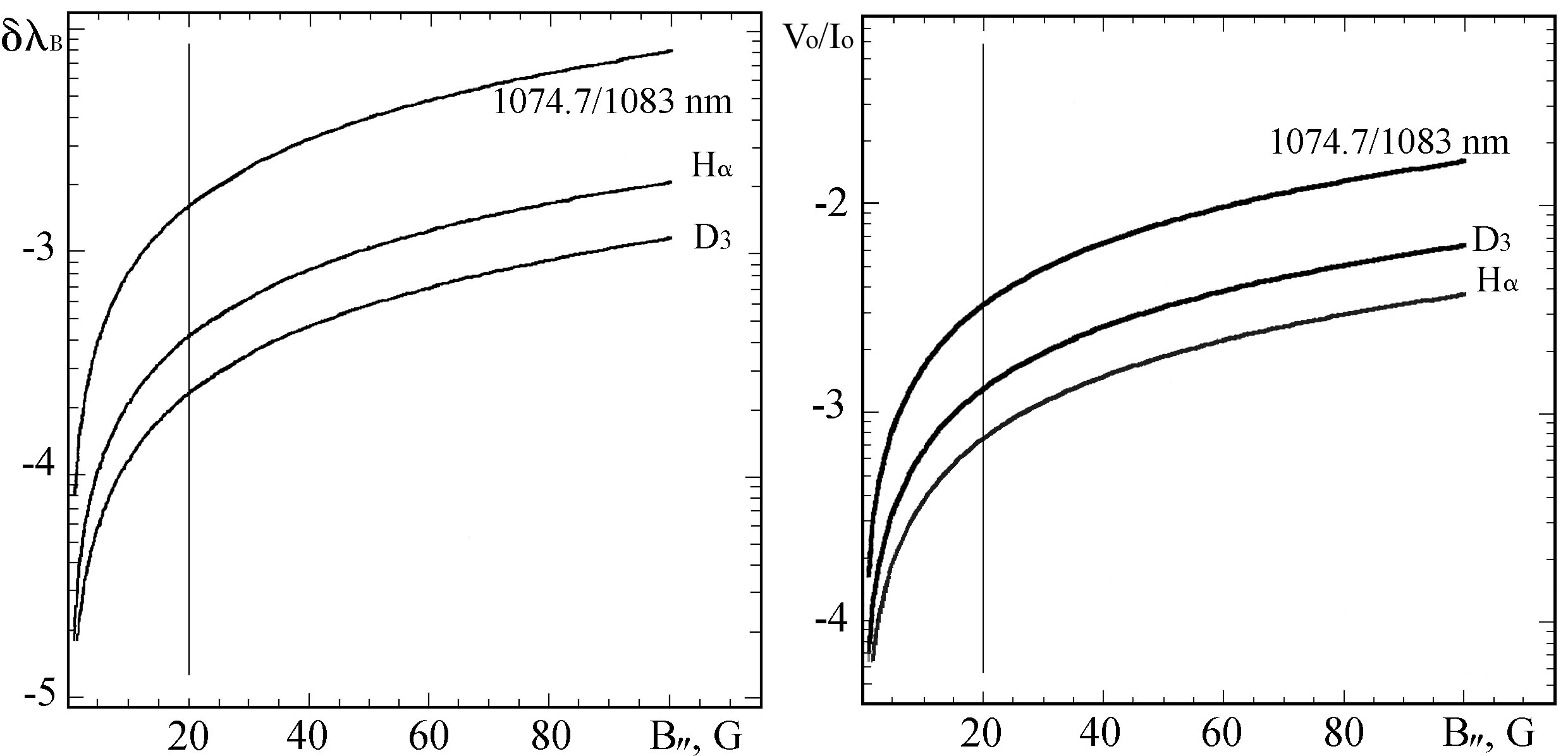}
}
              \caption{The Zeeman splitting
[$\delta\lambda_{B_{\prime\prime}}$] (left) and $k-$factor
[$V_{0}/I_{0}$] (right) versus longitudinal magnetic-field
strength [$B_{\prime\prime}$] for H {\tiny I} H$\alpha$, He {\tiny
I} D$_3$, and 1083.0 nm, and Fe {\tiny XIII} 1074.7 nm lines.}
   \label{Fig1}
   \end{figure}
%%%%%%%%%%%%%%%%%%%%%%%%%%%%%%%%%%%%%%%%%%%%%%%%%%%%%%%%%%%%fig1

In prominences, the chromosphere, and the corona $\Delta\lambda =
0.4$\,--\,1 \r{A}, and in the first approximation the upper part
of $I$-profiles is well fitted by a gaussian. In ``weak'' fields
the $V$-profile is proportional to the first derivative of the $I$
one.
\begin{equation}
V(\lambda)\approx \mathrm{dI} =
-I_0\frac{2(\lambda-\lambda_0)\times \mathrm{d}\lambda}{\Delta
\lambda_{D}^2}\exp^{-\left(\frac{\lambda-\lambda_{0}}{\Delta\lambda_D}
\right)^2},
\end{equation}
where $I_{0}$ is the peak of the line intensity, $\lambda_{0}$
denotes the wavelength of the emission line, $\lambda$ is measured
 from $\lambda_{0}$, and $\Delta\lambda_D$ is the Doppler width.
Equating the first derivative of $V$ to zero, we find the value of
the $V$ peak [$V_{0}$] and the wavelength corresponding to $V_{0}$
[$\lambda_{B}$] as follows:
\begin{equation}
V_{0}\approx
I_0\frac{\Delta\lambda_D\exp^{-\left(\frac{\Delta\lambda}{\sqrt{2}\Delta\lambda_{D}}\right)^2}}{\sqrt{2}\Delta\lambda_{D}^2}\delta\lambda_{B}\approx1.4\times
I_0\frac{\delta\lambda_B}{\Delta\lambda}; \quad
\lambda_{B}=\lambda_{0}\pm\frac{\Delta\lambda_{D}}{\sqrt{2}}\approx\pm0.71\Delta\lambda_{D}.
\end{equation}
Let us introduce a $k$-factor that is defined as the ratio of
$V_{0}$ to $I_{0}$ and is needed for further estimates. In other
words, $ k $ indicates the amplification factor in the $V$ channel
to record $I$ and $V$ on the same scale.
\begin{equation} k =
\frac{V_{0}}{I_{0}} \approx 1.4\times
\frac{\delta\lambda_B}{\Delta\lambda}.
\end{equation}
$k$($B_{\prime\prime}$) is shown in Figure 1 (right).
Sophisticated polarimeters were needed to record the $V$-profile
which is $7\times10^{-4}I$ when using H$\alpha$ line and
$B_{\prime\prime} = 20$ G.

\subsection{Feeding Optics of Previous Long-term Prominence
Magnetic-Field Measurements}

The direct magnetic-field determination in the upper solar
atmosphere is based on the circular and linear-polarization
analysis. The Zeeman analysis does not need any assumption on the
mechanism of radiation and allows an approach close to the limb.
The main steps of several long-term magnetic studies in
prominences are outlined  in the recent memoir by
\inlinecite{tandberg:2011}. The feeding optics used (telescopes
with the entrance aperture $< 0.5$ m) are briefly noted below.
Hereinafter, only references concerning the aspects of $I_{\rm
stray}$ are cited.
\begin{enumerate}
\item The first magnetic measurements in active prominences were made
by \inlinecite{zirin:1961} and \inlinecite{ioshpa:1962}:
Babcock-type type magnetographs, 30-cm solar tower telescopes,
B$_{\prime\prime}\approx100$ G.
\item Next successes were based on  magnetographs developed specifically for
Zeeman analysis in prominences \cite{lee:1965,lee:1969} and Climax
40-cm coronagraph: the magnetograph slit of
[$14-25^{\prime\prime}$]; an integration time up to 10 minutes.
\inlinecite{rust:1966} carried out magnetic research in quiescent
prominences (QP): $B_{\prime\prime}$ ranges from a few G to 10 G,
sometimes $20 - 30$ G. \inlinecite{harvey:1968},
\inlinecite{malville:1968}, and \inlinecite{harvey:1969} carried
out measurements in active prominences (AP): $B_{\prime\prime}=
40-200$ G, possible dependence on the phase of solar cycle, the
angle between the field vector and the long axis of prominences
[$\alpha] < 20^{\circ}$ \cite{tandberg:1970}.

\noindent Determinations of the magnetic-field vector in
prominences have been made by  \inlinecite{athay:1983} with the
advanced Stokes polarimeter and the 40-cm coronagraph of the
Sacramento Peak Observatory: the polarimeter slit of
$[7^{\prime\prime} \times 10^{\prime\prime}]$, an integration time
of two minutes.

\item  Contradictory results were reported by \inlinecite{smolkov:1971}
and \inlinecite{bashkirtsev:1971} for the first stage of their
measurements with the 50-cm horizontal solar telescope and the
magnetograph scanning across the line profile: the magnetograph
slit of $7^{\prime\prime}$, $B_{\prime\prime}$ up to 100 G in QP
and up to 1000 G in AP.

\item The next long-term Zeeman analysis was made with Nikolsky's
magnetograph developed in cooperation with Institute
d'Astrophysique de Paris
\cite{den:1977,nikolsky:1982,stepanov:1989} and the 50-cm domeless
refractive coronagraph: the magnetograph pinhole of
$8^{\prime\prime}$, an integration time of 30 seconds.

\noindent -- A piezo-scanning Fabry--Perot interferometer with a
pre-filter.

\noindent -- A LiNbO$_{3}$ crystal as an analyzer.

\noindent -- Measurements in the vicinity of the optical axis ($<
30^{\prime\prime}$).

\noindent -- The use of the magnetic-field etalon \cite{kim:2000}.

\noindent -- Compensation of the instrumental polarization
\cite{klepikov:1999}.

Results of the statistical analysis were as follows
\cite{kim:1990}: $B_{\prime\prime}$ in QP of several G, sometimes
reaching $30-40$ G; in AP $B_{\prime\prime} = 40-150$ G;
$\alpha\le25^{\circ}$; both the inverse and normal polarities may
exist in the same prominence.
\end{enumerate}
To summarize, only coronagraphs as feeding optics provided
long-term ``weak'' magnetic-field measurements [10\,--\,20 G] in
prominences.

\subsection{Non-object Signatures}
On average, $V_{0}$ depends on $I_{0}$, $\lambda$, $g$, and
$B_{\prime\prime}$. Nevertheless, the Zeeman diagnostics in
spicules are technically more complicated task despite the fact
that their intensities are greater than the intensity of bright
prominences. Significant noise appears when approaching the limb.

Non-object signatures complicate the direct near-limb Zeeman
diagnostics. Let $S/N$ be the signal-to-noise ratio. In our case
$S = V_{0}$ and $N$ is the noise in the $V$ channel caused mainly
by input of non-object signatures: $I_{\rm stray}$, $I_{\rm
cont}$, and $I_{\rm sky}$ that are one\,--\,three orders of
magnitude lower than $I$. Photon noise is assumed. In the first
approximation, $N \approx k\times\left[\sqrt{I_{\rm stray}} +
\sqrt{I_{\rm sky}} + \sqrt{I_{\rm cont}}\right]$. Let $S/N =
V_{0}/N$ be $\ge 3$. Using the expressions (2) and (3) we obtain
\begin{equation}
N \le 0.47\times I_0\frac{\delta\lambda_B}{\Delta\lambda} \quad
\Longrightarrow \quad \left[\sqrt{I_{\rm stray}} + \sqrt{I_{\rm
sky}} + \sqrt{I_{\rm cont}}\right] \le
0.47\frac{I_{0}\times\delta\lambda_B}{k\times\Delta\lambda}.
\end{equation}
Note that for reasonable non-object signatures (the total
$<10^{-5}$), $B_{\prime\prime} = 20$ G, $\delta\lambda_{B}$ and
$k$ derived from Figure 1, the above expression is satisfied for
$I_{0} > 7 \times 10^{-3}$ in H$\alpha$ and $>10^{-2}$ in IR lines
that correspond to prominences. The intensity of coronal lines is
much lower.  An increase of the integration time and the entrance
aperture is required to effectively increase the incoming flux.

\section{Reducing the Stray Light in Telescopes}
The existence of large-spread-angle stray light may significantly
affect polarization measurements. According to
\inlinecite{chae:1998}, the observed polarization degrees are
always underestimated. The main sources of ``parasitic''
background in the final focal plane of any telescope are the
following:
\begin{enumerate}
\item A ghost solar image produced by multiple
reflections in the primary lens.
\item Random inhomogeneities in the glass of the primary lens.
\item Departures of the surface of the primary optics
from a uniform shape.
\item Diffraction of the solar disk light at the entrance
aperture [$I_{\rm dif}]$.
\item Scattering at micro-roughness of the primary optics [$I_{\rm sc}$].
\item The sky brightness [$I_{\rm sky}$].
\item The continuum corona [$I_{\rm cont}$].
\item The dust on the primary optics [$I_{\rm dust}$].
\end{enumerate}

Reducing items \textit{i})\,--\textit{iv}) was partly discussed by
\inlinecite{newkirk:1963}. Items \textit{iv})\,--\textit{vi}) are
negligible during total solar eclipse (TSE), and they dominate for
high-altitude coronagraphic observations. Reducing the item
\textit{viii}) for 53-cm refractive primary optics (our
experience) was made by cleaning the lens before each set of
observations. But this item becomes very important for
large-aperture reflective primary optics.  Below the role of the
stray light for magnetic-field measurements will be discussed.
Hereinafter we denote $I_{\rm stray}$ as $I_{\rm dif}+I_{\rm sc}$.

\subsection {Reducing the Stray Light Caused by Diffraction of the
Bright Round Source at the Round Aperture}

The diffraction of the bright round source at the round aperture
was treated by \inlinecite{nagaoka:1920}. For estimations of
$I_{\rm dif}$, we used the simplified expression suggested by
\inlinecite{sazanov:1968} which is valid in the range R $<1.3$
R$_\odot$. Deviations not more than $20\,\%$ are expected as
compared with values based on Nagaoka's equations.
\begin{equation}
\log{I_{\rm dif}(\rho)}=\log\frac{2\lambda}{\pi^3D\gamma_0} +
\log\left(\frac{\sqrt{1+(\nu-1)^2}}{\nu-1}-\frac{\sqrt{1+(\nu+1)^2}}{\nu+1}\right)
- 0.27(\nu - 1) - 0.017,
\end{equation}
where $\rho$ is the distance from the solar disk center in the
units of R$_\odot$,  $\nu = \frac{\rho}{a}$, $a = \frac{\pi
D}{\lambda}\gamma_0$ is the radius of the round source in
arbitrary units, $D$ is the diameter of the primary lens,
$\gamma_0$ is the angular radius of the source. $I_{\rm dif}$(R)
is shown in Figure 2 for H$\alpha$ (left) and near IR (right)
lines for 0.5 m, 1.6 and 4 m apertures. Vertical lines indicate
typical maximum altitudes observed [$h$] of QP ($40^{\prime\prime}
\approx0.042$ R$_{\odot}$), active region filaments (ARF)
($25^{\prime\prime}\approx0.026$ R$_{\odot}$), and the upper
chromosphere ($4^{\prime\prime}\approx0.004$ R$_{\odot}$).
Hereinafter intensities, brightness, equivalent width are given in
units of the 1 \r{A} nearby solar-disk continuum.

\begin{figure}    %%%%%%%%%%%%%%%%%% FIGURE 2
\centerline{\includegraphics{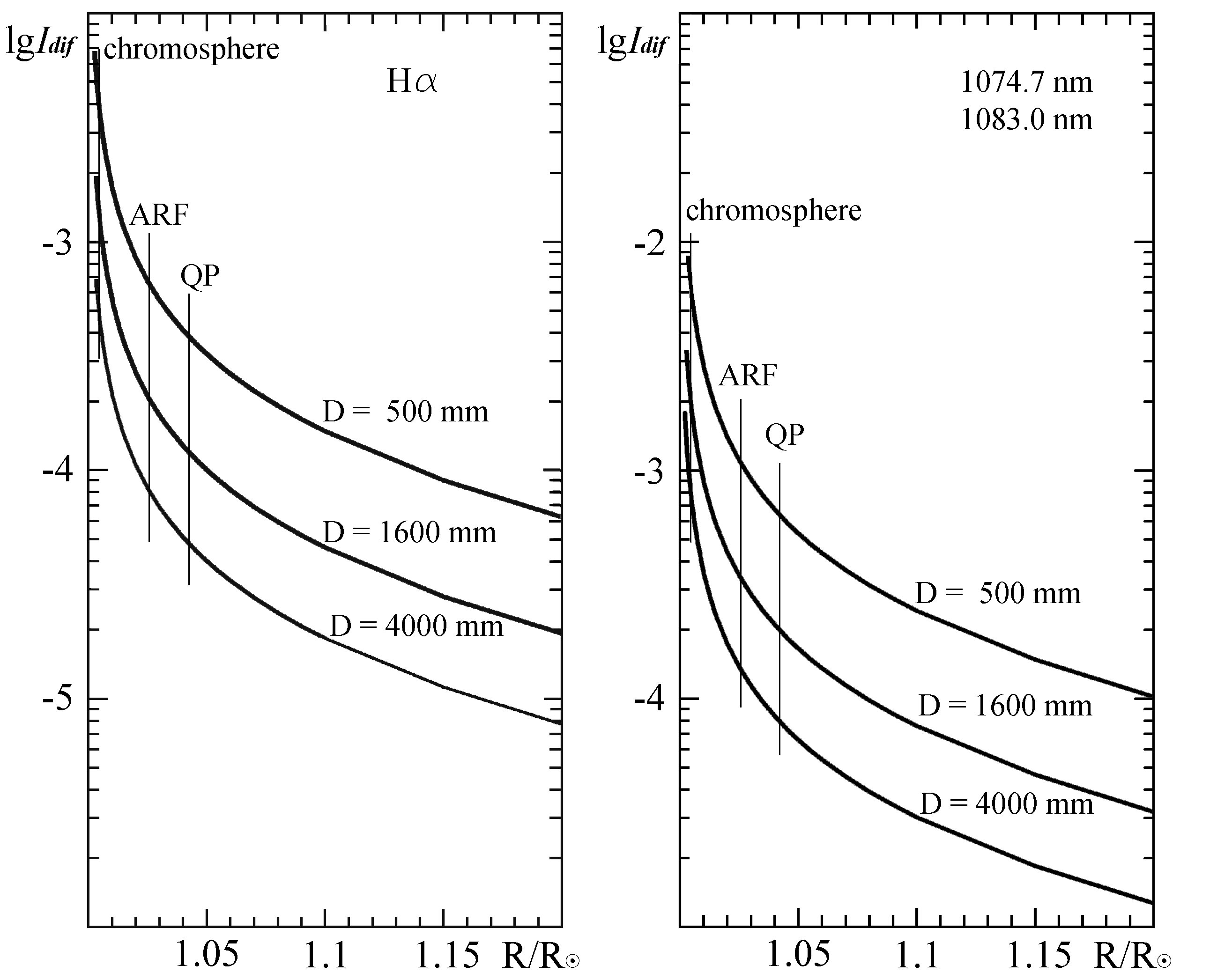}}
\caption{Stray light caused by diffraction of the solar disk light
at the entrance aperture. Left: H$\alpha$ line. Right: the He
{\tiny I} 1083.0 nm and Fe {\tiny XIII} 1074.7 nm lines.}
\label{Fig2}
\end{figure}%%%%%%%%%%%%%%%%%%%%%%%%%%%%%%%%%fig2

It is seen that the stray light in H$\alpha$ caused by diffraction
at the 0.5 m aperture of a conventional (non-coronagraphic)
telescope can reach $10^{-3}$ at prominence heights and is $>
5\times 10^{-3}$ at the chromosphere level.

\subsubsection{Coronagraphic Technique (the Lyot Method)}
The coronagraphic technique suggested by \inlinecite{lyot:1931} is
based on the masking in the primary focal plane and in the plane
of the exit pupil to eliminate \textit{i}) and \textit{ii}), and
to minimize the input of \textit{iv}). The optical sketch of the
Lyot-type coronagraph has the primary single lens, the primary
focal plane, the mask in the primary focal plane (an artificial
Moon), the field lens, the mask in the plane of the exit pupil
(the Lyot stop), the relay optics, and the final focal plane.
Multiple Fresnel reflections at the surfaces of the primary lens
create a system of the solar-disk images, decreasing in
brightness. For a single primary lens with the refraction index $n
= 1.5$, the brightness ratio of the first, most bright reflection
to the the solar disk one is $(n-1)^{4}/(n+1)^{4} = 1/625$, and
the ratio of the primary focal length to the space between the
image and the lens is  $[2/(n-1)] +3 = 7$. A round screen in the
center of the Lyot stop results in removal of the reflection. The
procedure is not applicable for the multi-lens primary optics, as
the brightest reflection is near the primary focal plane. The
correct use of the Lyot method results in $I_{\rm dif}$ reducing
in one to two orders of magnitude depending on the size of the
mask in the primary focal plane and in the plane of the Lyot stop.
In practice, reducing $I_{\rm stray}$ by $\approx50-100$ (the
coronagraphic efficiency [$K$]) for prominences and 10\,--\,25 for
the chromosphere can be achieved depending on the height observed.

\subsubsection{Multiple Cascade Coronagraphic Technique}
To our knowledge, the multiple-cascade coronagraphic technique has
not been used practically. \inlinecite{terrile:1989} found that an
additional factor of more than ten can be achieved through
multiple stages of apodizing in both the focal plane and in the
Lyot-stop plane. The calculated point spread function (PSF) showed
that in such a hybrid coronagraph $I_{\rm dif}$ can be reduced by
more than three orders of magnitude.

We used the two stage coronagraphic approach for the last version
of Nikolsky's magnetograph \cite{stepanov:1989}. The main goal was
to match the focal ratio of the coronagraph with the spectral
resolution of the Fabry--Perot interferometer through the
inclusion of an additional focal and the Lyot-stop planes. This
complicated the optical adjustment of the ``coronagraph +
magnetograph'' assembly. Depending on the brightness of
prominences, the magnetic-field strength, and the height observed,
the signal-to-noise ratio became two\,--\,three times better.

\subsubsection{Apodizing with a Special Mask in the Plane of an Entrance Aperture}

The diffraction pattern in the focal plane is the result of
discontinuity of the transmission function [$G$] (or its
derivatives) of the entrance aperture. The characteristic
frequency of the damping intensity oscillations depends on the
distance between the points of discontinuity, the asymptotic
damping rate depends on the order of the derivative in which the
continuity of $G$ ($G^{2}$ in intensity) is broken. For a round
aperture, there is a discontinuity in the zero derivative:
$G(\rho) = 1$ in the range $\rho < 1$ and $G = 0$ in the range
$\rho \ge 1$, where $\rho$ is the distance from the center of
aperture. In the case of a point source, it creates the Airy
diffraction pattern $I \propto [J_1(r)/r]^{2}$ with intensity
damping as $r^{-3}$, where $J_1(x)$ is the Bessel function of the
first kind. A mask with variable transmission
[$G(\rho)=1-\rho^{2}$] has discontinuities only in the first
derivative. The use of such a mask in the plane of the entrance
aperture results in the diffraction image $I \propto
{\{[J_1(r)+J_3(r)]/r\}}^{2}$ with the more effective damping as
$r^{-5}$.

We considered an extended object, \textit{e.g.}, the Sun
\cite{kim:1995} using Nagaoka's equations \cite{nagaoka:1920}. The
stray light in the center of the solar disk image caused by
diffraction at the entrance aperture  is given by
\begin{equation}
I_{\rm dif}(0)=1-J_0^{2}(a) - J_1^{2}(a),
\end{equation}
where $a=(\pi D\sin{\gamma_0})/\lambda, \lambda$ is the
wavelength, $\gamma_0$ is the angular radius of solar disk, and
$D$ is the diameter of aperture. If $a\gg1$, then $I_{\rm
dif}(0)\approx1-2/(\pi a)$. For $\gamma_0=960^{\prime\prime}$,
$\lambda=600$ nm, $D=200$ mm, we obtain $a=4874$ and $I_{\rm
dif}(0)=(1-1.3)\times 10^{-4}$.

In the case when the entrance aperture is apodized by the mask
$G=(1-\rho^{2})$, we found the relation between the apodized
[$I^{a}_{\rm dif}$] and non-apodized [$I_{\rm dif}$] for points at
the angular distance $\gamma = (1+\epsilon)\gamma_0$ from the disk
center:
\begin{equation}
I^{a}_{\rm dif}(\epsilon)\simeq 2/3\pi^{4}{[I_{\rm
dif}(\epsilon)]}^{3}.
\end{equation}
Let us estimate the efficiency of the mask for chromospheric and
prominence heights and $a\approx5000$.
\begin{itemize}
\renewcommand{\labelitemi}{$-$}
\item The upper chromosphere heights: $h = 4^{\prime\prime}$, $\epsilon=0.004$
$(\epsilon\gamma_0 = 4^{\prime\prime})$. Then $I(0.004)=10^{-2}$
and $I^{a}(0.004)=0.7\times 10^{-4}$. Calculated efficiency up to
$10^{2}$ can be achieved.
\item Quiescent prominence heights: $h = 40^{\prime\prime}$, $\epsilon =
0.04$  ($\epsilon\gamma_0=40^{\prime\prime})$. Then $I_{\rm
dif}(0.04)=10^{-3}$ and $I^{(a)}_{\rm dif}(0.04)=0.7\times
10^{-7}$. Calculated efficiency up to $10^{4}$ can be achieved.
\end{itemize}
No classical Lyot-type coronagraphs are needed. Note that the mask
$(1-\rho^{2})$ reduces transmittance by a factor of three.

\subsection{Comments on Scattering by Micro-Roughness of
the Primary Optics}

In this subsection, we do not analyze scattering by
micro-roughness of the primary optics. This is a topic requiring
detailed studies. In the case of the statistical nature of the
micro-roughness, $I_{\rm sc}$ is proportional to the square of the
average height of the inhomogeneity (RMS). The fabrication of
super-smooth primary optics is the key technology for creating a
low-scattered-light coronagraph. Reflecting optics are achromatic
and do not depend on bulk inhomogeneities of the material compared
to a refractor. Let $n$ be the index of refraction. At the same
value of RMS, the energy scattered by a reflecting surface ($n^{*}
= -1$) [$I_{\rm sc}$]  is greater by a factor of
$[(n^{*}-1)/(n-1)]^{2}=16$ compared to the refractive case.
Possible ways to reduce the scattered light include the following.
\begin{itemize}
\renewcommand{\labelitemi}{$-$}
\item The use of a super-smooth primary optics with RMS = 3\,--\,10
\r{A}. Pioneering studies performed by \inlinecite{socker:1988}
showed that the $I_{\rm stray}$ of a 9.8 cm diameter super-smooth
silicon mirror is comparable with the stray light of a single
lens.
\item The use of moderately smooth primary optics with a given
profile of the micro-relief can significantly reduce the scattered
light in the range of interest. According to numerical
calculations by \inlinecite{romanov:1991} made for the
Earth-environment monitoring, RMS of 25 \r{A} and the spatial
period of the micro-relief (the correlation length of
inhomogeneities) of $0.76-0.78$ $\mu$ can provide scattered light
of $\approx10^{-5}$ in the range $<1.1$ R$_{\oplus}$ where
R$_{\oplus}$ is the radius of the Earth.
\end{itemize}

\section{Acceptable Level of the Stray Light in
Telescopes for Zeeman Diagnostics}

Using Equation (4), Figures 1 and 2, let us estimate the
acceptable level of the stray light in telescopes for Zeeman
diagnostics with the signal-to-noise-ratio $\ge3$. Several
conditions exist.
\begin{itemize}
\renewcommand{\labelitemi}{$-$}
\item The stray light is caused by diffraction at the entrance aperture
(a super-smooth primary optics).
\item The noise of the recording assembly is negligible.
\item $W$ is the equivalent width of the emission line.
\item The instrumental width is $0.7 \Delta\lambda_{D}$
 $\approx0.5 \Delta\lambda$ to achieve the maximum
signal-to-noise ratio \cite{nikolsky:1985}.
\item In the case of scattering by aerosols ($\sim\lambda^{-2}$), $I_{\rm sky} =
10^{-5}$ in H$\alpha$ and $I_{\rm sky}\approx 2\times 10^{-6}$ in
the IR lines.
\item $I_{\rm cont}$ passed through the instrumental profile is $< 10^{-6}$.
\end{itemize}

\textit{H$\alpha$ bright prominences, B$_{\prime\prime} = 100$ G}:
W $= 10^{-1}$, $\Delta\lambda \approx 0.6$ \r{A}. $I_{0} = 5\times
10^{-2}$, $\delta\lambda_{B}=2\times 10^{-3}$ and $k = 4\times
10^{-3}$ (Figure 1). The required $I_{\rm stray}$ should be $\le
5\times 10^{-4}$. Referring to Figure 2 (left), we see that the
aperture of 0.5 meter provides this level of scattered light at $h
> 35^{\prime\prime}$. The first magnetic measurements in
prominences with 30 cm solar tower telescopes as feeding optics
confirm this \cite{zirin:1961,ioshpa:1962}.

A non-coronagraphic 4 meter telescope with a super-smooth primary
optics can provide Zeeman diagnostics of 100 G field strengths in
bright prominences from $h\ge 10^{\prime\prime}$ as well.

\textit{H$\alpha$ moderate brightness prominences,
B$_{\prime\prime}= 20$ G}: $W = 2\times 10^{-2}$, $\Delta\lambda
\approx 0.6$ \r{A}. $I_{0} = 10^{-2}$, $\delta\lambda_{B} =
4\times 10^{-4}$ and $k = 7\times 10^{-4}$ (Figure 1). The
required $I_{\rm stray}$ should be $\le 2 \times 10^{-5}$. Only
coronagraphs (at the limits of 0.5 meter, and reliably with 4
meter apertures) provide the required $I_{\rm stray}$ from heights
$>20^{\prime\prime}$. Note the crucial role of the sky brightness,
the polarimeter performance, and the integration time.

\textit{Limb chromosphere, He {\footnotesize I} 1083.0 nm,
B$_{\prime\prime}= 20$ G}: $W \approx 3\times 10^{-2}$,
$\Delta\lambda \approx 0.8$ \r{A}. $h = 4^{\prime\prime}$, $I_{0}
= 1.5\times 10^{-2}$, $\delta\lambda_{B} = 1.3\times 10^{-3}$ and
$k = 3\times 10^{-3}$ (Figure 1). Then the required $I_{\rm stray}
\le 1.6\times 10^{-5}$ will be provided with the 4 meter
coronagraph in which the scattered light is reduced by $30-40$.

\textit{Corona, Fe {\footnotesize XIII} 1074.7 nm,
B$_{\prime\prime} = 20$ G}: $W = 2 \times 10^{-4}, \Delta\lambda
\approx 0.8$ \r{A}. The expected total of all non-corona
signatures is $\approx 10^{-5}$. $\delta\lambda_{B} = 1.5 \times
10^{-3}$, and $k = 3.3 \times 10^{-3}$ (Figure 1). $I_{0} =
10^{-4}$, that is increasing $I_{0}$ up to $10^{2}$ (see above) is
required. A 4 meter aperture allows an increase in the flux by 64
times as compared to a 0.5 meter one.  In this case, the
conditions will be similar to magnetic-fields measurements in
moderate brightness prominences when using a 0.5 meter
coronagraph. $I_{\rm sky}$, the performance of the polarimeter,
and the integration time become important factors.

\section{Reduction of Random and Systematic Errors}

We have developed an approach for high-precision linear
polarimetry with actual accuracy of $2\%$ for the linear
polarization degree [$p$] and $2^{\circ}$ for the polarization
angle [$\chi$] that allow us to obtain ``polarization images'' of
an object: 2D distributions of $p$, $\chi$, and the sign of
$\chi$. The description of the last version was presented recently
\cite{kim:2011a}. The key components are the following:
\begin{enumerate}
\item Low level of the sky brightness, $I_{\rm sky}$.
\item A low level of the stray light in telescopes, $I_{\rm stray}$.
\item Uniformity of the polarizer performance for any
``point'' of the image.
\item Reduction of random errors based on ``statistics'': the use of
24 orientations of a polarizer instead of traditional three.
\item Reduction of systematic errors based on Stokes-vector
presentation of the light and the solution of the over-determined
system of 24 equations (the number of equations is greater than
the the number of unknowns) by the least squares.
\end{enumerate}
The role of the last two points is shown below. We use the total
solar eclipse of 29 March 2006 observations, as $I_{\rm sky}$ and
$I_{\rm stray}$  were negligible during totality. The approach was
applied to the red polarization movie of the continuum corona to
check the potential of our method, the reliability of the
predicted accuracy, and the importance of items \textit{iv}) and
\textit{v}) for the near-limb polarimetry. The red series of 24
sequential frames centered at 25 seconds before the third contact
[T$_3$] was treated in 2 ways to search for evidence for H$\alpha$
prominences in polarization as our red filter transmitted
H$\alpha$ line. Until now, measurements of linear polarization in
low-brightness prominences [$I < 10^{-2}$] are rarely carried out
in spite of available advanced polarimeters. Figure 3 shows the 2D
distribution of $p$ in the range $0-8\,\%$ for distances $<1.1$
R$_\odot$ above the SW limb. The left distribution is based on
three orientations of a polarizer spaced by $120^{\circ}$ and
exhibits a noise of $3-7\,\%$. The right one is based on 24
orientations of the same series for the same position angle range
and clearly reveals the polarized feature at position angle $P =
251^{\circ}$ with $p = 3-7\,\%$. Position angles are measured
counter-clockwise from the N Pole of the Sun. The solar and lunar
limbs, the scales of heights and $p$, and  the N direction are
shown. The step of the $p$-scale is $2\,\%$. Synoptic data from
the Pulkovo Observatory identify this $p$-feature with a low
brightness H$\alpha$ prominence ($10^{-3}-10^{-2}$) at the same $P
= 251^{\circ}$. It is known that in the absence of longitudinal
magnetic fields, the polarization degree in prominences increases
from $3$ to $8\,\%$ with an increase in height from
$10^{\prime\prime}$ to $96^{\prime\prime}$. The distribution
agrees with the expected $p$-values and indirectly confirms the
accuracy $<2 \,\%$. We note that these ``raw'' 2D distributions
were obtained only to test the ability of our method to
distinguish the near-limb several-percent linear polarization and
to verify the accuracy $< 2 \,\%$. Actual values of $p$ are
expected to be $1-2\,\%$ lower as the transfer to intensities was
based on a polynomial approximation of degree four over the wide
range of densities from the background to prominences and no
corrections for the red coronal continuum input was made. The
corrected eclipse linear polarimetry in prominences as well as
$p$, $\chi$, and the sign of $\chi$ images will be discussed in a
separate article.
\begin{figure}    %%%%%%%%%%%%%%%%%% FIGURE 3
\centerline{\includegraphics{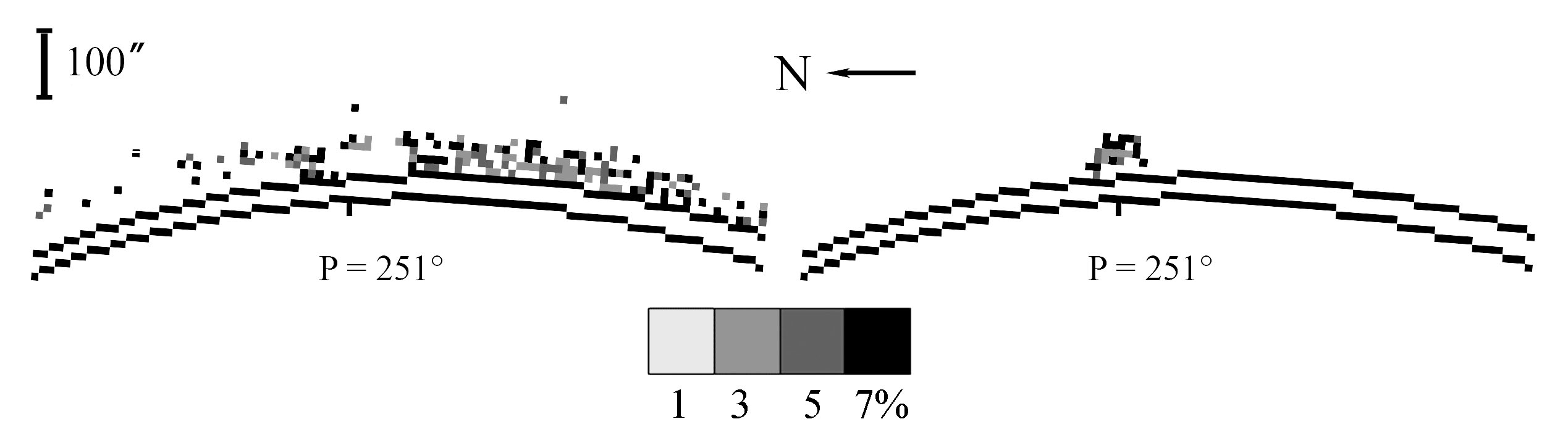}}
\caption{The 2D distribution of $p$ in the red spectral interval
including H$\alpha$ line based on three orientations of a
polarizer (left) and on 24 orientations (right). The right
$p$-image reveals the polarized feature at $P = 251^{\circ}$,
which is identified with the low-brightness H$\alpha$ prominence.}
\label{Fig3}
\end{figure}%%%%%%%%%%%%%%%%%%%%%%%%%%%%%%%%%fig3

\section{Summary}
The stray light [$I_{\rm stray}$] seems to be a crucial factor
determining the reliable near-limb $V$-profile recording in the
range $<1.2$ R$_{\odot}$. Our estimates of non-object signatures
quantitatively show that the most advanced polarimeter will be
powerless if $I_{\rm stray}$ in a feeding optics exceeds the
acceptable level. A brief comparison of several ways of $I_{\rm
stray}$ reduction results in the following.
\begin{itemize}
\renewcommand{\labelitemi}{$-$}
\item The well-known coronagraphic technique
(the Lyot method) provides the coronagraphic efficiency [$K$] of
10\,--\,100 depending on the size of mask in the primary focal
plane and in the plane of the Lyot stop (depending on the object
under study).
\item According to our experience, multiple stages of apodizing
in both the focal plane and in the Lyot-stop plane provides $K =
2-3$ and complicates the optical adjustment of the ``coronagraph +
magnetograph'' assembly.
\item Apodizing with a mask with variable transmission placed in
the plane of an entrance aperture [$G(\rho)=1-\rho^{2}$], where
$\rho$ is the distance from the center of aperture. For a 200 mm
aperture, the calculated efficiency factor up to $10^{2}$ at the
upper chromosphere level [$4^{\prime\prime}$] and up to $10^{4}$
at the QP heights [$40^{\prime\prime}$] can be achieved. Recent
technology advances allow the discussion on manufacturing such a
mask. Synoptic chromospheric and prominence magnetic research seem
to be reliable.
\item The use of a super-smooth primary optic with RMS = 3\,--\,10
\r{A} or a moderately smooth primary optics (RMS of 25 \r{A}) with
a given profile of the micro-relief can significantly reduce the
scattered light in the range of interest.
\item The important role of reduction of random and systematic
errors is shown by the example of eclipse linear polarimetry of
prominences.
\end{itemize}
Estimation of the acceptable level of the stray light in 0.5, 1.6,
and 4 meter aperture telescope for Zeeman diagnostics with the
signal-to-noise-ratio of $>3$ show that the 4 meter-aperture ATST
with $I_{\rm stray} < 10^{-5}$ will provide $V$-profile recording
of ``weak'' fields [$B_{\prime\prime} = 10-20$ G] in prominences
in visual and IR lines, the limb chromosphere and corona in IR
lines with the finest magnetic ``resolution'' comparable with the
characteristic size of the structures.

%%%%%%%%%%%%%%%%%%%%%%%%%%%%%%%%%%%%%%%%%%%%%%%%%%%%%%%%%%%%%%%%%%%%%%%%%%%
\begin{acks}
This work was partially supported by research project No
11-02-00631 of Russian Foundation for Basic Research. We are very
indebted to the referee for corrections and important comments
concerning the sources of stray light.
\end{acks}

%%% BIBLIOGRAPHY %%%%%%%%%%%%%%%%%%%%%%%%%%%%%%%%%%%%%%%%%%%%%%%%%%%%%%%%%%%
\bibliographystyle{spr-mp-sola}
\bibliography{sp_kim}

\IfFileExists{\jobname.bbl}{} {\typeout{}
\typeout{****************************************************}
\typeout{****************************************************}
\typeout{** Please run "bibtex \jobname" to obtain} \typeout{**
the bibliography and then re-run LaTeX} \typeout{** twice to fix
the references !}
\typeout{****************************************************}
\typeout{****************************************************}
\typeout{}}
\end{article}
\end{document}